\documentstyle[epsf]{mn}
\newcommand{\etal}{{et al}\/.}

\begin{document}
\title[Magnetic fields in 3C\,33 and 3C\,111]{Magnetic field strengths
in the hot spots of 3C\,33 and 3C\,111}
\author[M.J.~Hardcastle \etal]{M.J. Hardcastle$^1$,
M. Birkinshaw$^{1,2}$ and D.M. Worrall$^{1,2}$\\
$^1$Department of Physics, University of Bristol, Tyndall Avenue,
Bristol BS8 1TL\\
$^2$Harvard-Smithsonian Center for Astrophysics, 60 Garden Street,
Cambridge, MA 02138, U.S.A.}
\maketitle
\begin{abstract}
We report on {\it ROSAT} HRI observations of the nearby powerful radio
galaxies 3C\,33 and 3C\,111, which both have detected optical hot
spots. We find nuclear X-ray sources in both objects, but no X-ray
emission from the hot spots. This confirms the presence of a
high-energy cutoff in the spectrum of synchrotron-emitting
electrons. Since these electrons necessarily scatter the synchrotron
photons by the inverse-Compton process, our upper limits on the X-ray
fluxes of the hot spots allow us to set lower limits of a few
nanotesla on their magnetic flux density, close to or greater than
the fields implied by equipartition of energy
between radiating particles and magnetic field.
\end{abstract}

\section{Introduction}

Few good methods exist of measuring the magnetic fields in the
extended components of extragalactic radio sources. Discussion of the
energetics and dynamics of radio sources is often based on the
assumption that there is equipartition of energy between magnetic
fields and energetic particles (or, effectively equivalently, that the
total energy has the minimum value consistent with the production of
the observed synchrotron radiation) but it has seldom been possible to
test this, and there is no strong {\it a priori} reason why it should
be true (e.g.\ Leahy 1990). X-ray observations of radio hot spots
provide one good way of making a relatively model-independent
measurement of the magnetic field, but such a measurement has only
been performed in one source, Cygnus A (Harris, Carilli \& Perley
1994) so far. In this paper we discuss deep observations with the {\it
ROSAT} HRI of two further radio galaxies. These objects have optically
detected hot spots, and so are ideal targets for investigation of the
magnetic field.

Radio sources with optical hot spots are of particular interest in the
X-ray, because the hot spot electron energy spectrum is well
constrained if it is assumed that the optical emission is synchrotron
radiation from the population of electrons responsible for the radio
emission. Meisenheimer \etal\ (1989) have analysed the
radio-to-optical spectra of a number of these sources.  With the
exception of Pictor A, all the sources they study show a spectral
turnoff in or around the optical regime, corresponding in a simple
synchrotron model to an upper energy cutoff of $\sim 2 \times 10^{11}$
eV. The subjects of this paper, 3C\,33 and 3C\,111, are both sources
with optical hot spots discussed in Meisenheimer \etal\ (1989). The
radio-optical synchrotron spectrum in these objects is therefore well
known.

A variety of possible mechanisms exists for X-ray emission from radio
hot spots. Synchrotron emission in the X-ray is possible, and has been
suggested as an explanation for the X-ray emission from the jet of M87
(Biretta, Stern \& Harris 1991) and from the hot spot of 3C\,390.3
(Prieto 1997). However, objects with high-frequency cutoffs in the
optical, as inferred by Meisenheimer \etal\ for 3C\,33 and 3C\,111,
are unlikely based on spectral extrapolation to emit significant
synchrotron radiation at X-ray frequencies. Recently {\it ROSAT} data
(Harris \etal\ 1994) have been used to show that the hot
spots of Cygnus~A, undetected at optical wavelengths, radiate
significantly in the X-ray. The X-ray flux is well above that
predicted by a purely synchrotron model, given the inferred
high-energy cutoff in the electron spectrum, and Harris \etal\
interpret this as evidence that hot spots emit X-rays by a synchrotron
self-Compton process (SSC: inverse Compton up-scattering of the
synchrotron photons by the energetic electrons), although more exotic
emission mechanisms such as a proton-induced cascade (PIC; Mannheim,
Biermann \& Kruells 1991) are not ruled out by the observations.  In
addition, inverse-Compton scattering of photons from the cosmic
background radiation (CBR) and from the radiation generated by the
host galaxy and the central engine is an obligatory process.

Because the emissivity from any inverse-Compton process depends upon
the number density of electrons, it is possible to use observations of
this process to constrain the magnetic field in the scattering
regions, provided the synchrotron spectrum and geometry are well
known. This allows us to test the equipartition assumption. The
observations of Harris \etal\ (1994) were consistent with
equipartition, but it is important to test whether this applies to
radio sources in general. The high resolution and sensitivity of {\it
ROSAT} makes it the first X-ray observatory likely to detect and
distinguish inverse-Compton emission from hot spots.

3C\,33 and 3C\,111 are both FRII (Fanaroff \& Riley 1974) radio
galaxies. Their redshifts and radio properties are summarized in Table
\ref{sources}. Both are well studied in the radio (for 3C\,33 see
Dreher 1981, Rudnick 1988, 1989, Rudnick \& Anderson 1990, Leahy \&
Perley 1991: for 3C\,111 see Leahy \& Williams 1984,
Linfield \& Perley 1984, Leahy \etal\ 1997).  The southern hot spot of
3C\,33 (Meisenheimer \& R\"oser 1986; Meisenheimer \etal\ 1989; Crane
\& Stiavelli 1992) and the northern hot spot of 3C\,111 (Meisenheimer
\etal\ 1989) are optical synchrotron sources. 3C\,33 is a narrow-line
radio galaxy, while 3C\,111 is a broad-line object often classed as a
Seyfert 1. Both appear isolated on optical sky-survey plates.

$H_0 = 50$ km s$^{-1}$ Mpc$^{-1}$ and $q_0 = 0$ are assumed
throughout the paper.

\section{The data}

\subsection{3C33}

3C\,33 was observed with the {\it ROSAT} HRI for a total of 51.9~ks
between 1996 January 16 and 22. We analysed the data using the Post
Reduction Off-line Software (PROS), filtering out time intervals with
a `maximum high background' level of $> 5 \times 10^{-7}$ counts
pixel$^{-1}$ s$^{-1}$ so as to maximise our chances of detecting faint
components. This process left us with 47.7 ks of good data.

\begin{figure*}
\begin{center}
\leavevmode
\epsfysize 13cm
\epsfbox{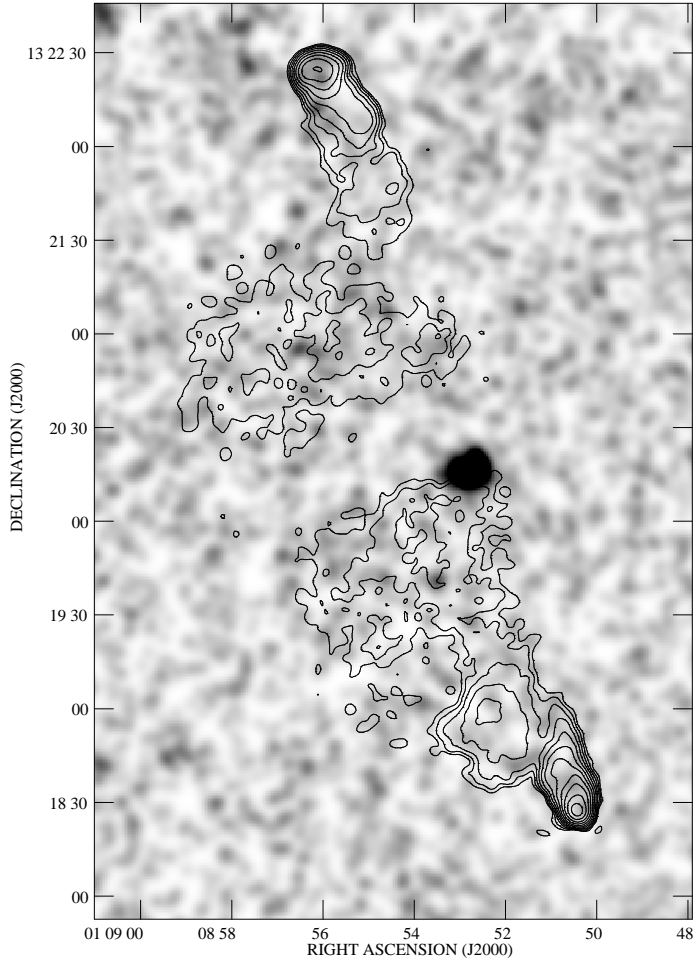}
\caption{1.5-GHz radio contours of 3C\,33 at 4.0-arcsec resolution
(from Leahy \& Perley 1991) overlaid on the HRI X-ray image, smoothed
with a Gaussian with $\sigma = 4$ arcsec. Radio contours at $2 \times
(1,2,4\dots)$ mJy beam$^{-1}$.}
\label{3C33-pic}
\end{center}
\end{figure*}

Fig.\ \ref{3C33-pic} shows an overlay of the radio contours of Leahy
\& Perley (1991) on a
smoothed X-ray image. An X-ray source is detected at a J2000.0
position of 01 08 52.7 +13 20 15.8, roughly 3 arcsec from the radio
core position on the radio map. We identify this with emission
from the nuclear regions or the host galaxy of 3C\,33, since
misalignments of up to 10 arcsec are expected with the absolute
position errors of {\it ROSAT}. Measuring the counts in a circle of
radius 1 arcmin about the centroid and taking background measurements
from a concentric annulus between 1 and 2 arcmin, we find a count rate
for this source of $\sim 3 \times 10^{-3}$ counts s$^{-1}$,
corresponding to a total of $140 \pm 30$ counts. The equivalent {\it
Einstein} IPC count rate would be $\sim 5 \times 10^{-3}$ counts
s$^{-1}$, assuming a power-law spectrum with photon index 1.8 and
galactic absorption. This is consistent
with the upper limit of $9 \times 10^{-3}$ counts s$^{-1}$ found by
Fabbiano \etal\ (1984) using the IPC.

Radial profile fitting in the region described
above, using 7 logarithmically spaced bins, suggests that it is
reasonably well modelled as a point source with little or no extended
structure. Following Birkinshaw \& Worrall (1993), we obtain a
model-independent upper limit on the source size by fitting the radial
profile with a broadened PSF; the best-fit FWHM for the source is then
6 arcsec and we find an upper limit on the FWHM of 9 arcsec at the 99
per cent confidence level. As errors in the {\it ROSAT} aspect
solution can introduce spurious extension of up to about 10 arcsec (e.g.\
Morse 1994), these results are consistent with the source being
unresolved. The upper limit then implies with a high degree of confidence
that the X-ray emission comes from a region less than 14 kpc in
size.

The lack of spectral information on this component introduces an
uncertainty in the determination of luminosity; further, if it is
emission from regions near to the AGN it may be
obscured (cf.\ Allen \& Fabian 1992), since 3C33 is a narrow-line
object, leading us to underestimate its true luminosity. With only
galactic absorption ($N_H = 3.9 \times 10^{20}$ cm$^{-2}$, Stark
\etal\ 1992) the 0.1--2.4 keV luminosity of the component is $5.6
\times 10^{42}$ ergs s$^{-1}$, if a power-law spectrum with photon
index 1.8 is assumed. If the intrinsic absorbing column were
comparable to that inferred for the power-law component of Cygnus A
($\sim 10^{23}$ cm$^{-2}$; Arnaud \etal\ 1987, Ueno \etal\ 1994) then
the true luminosity would be $\sim 10^{46}$ ergs s$^{-1}$, which would
be extraordinarily high for an object of this moderate radio
power.

\begin{figure*}
\begin{center}
\leavevmode
\hbox{\epsfysize 10.5cm
\epsfbox{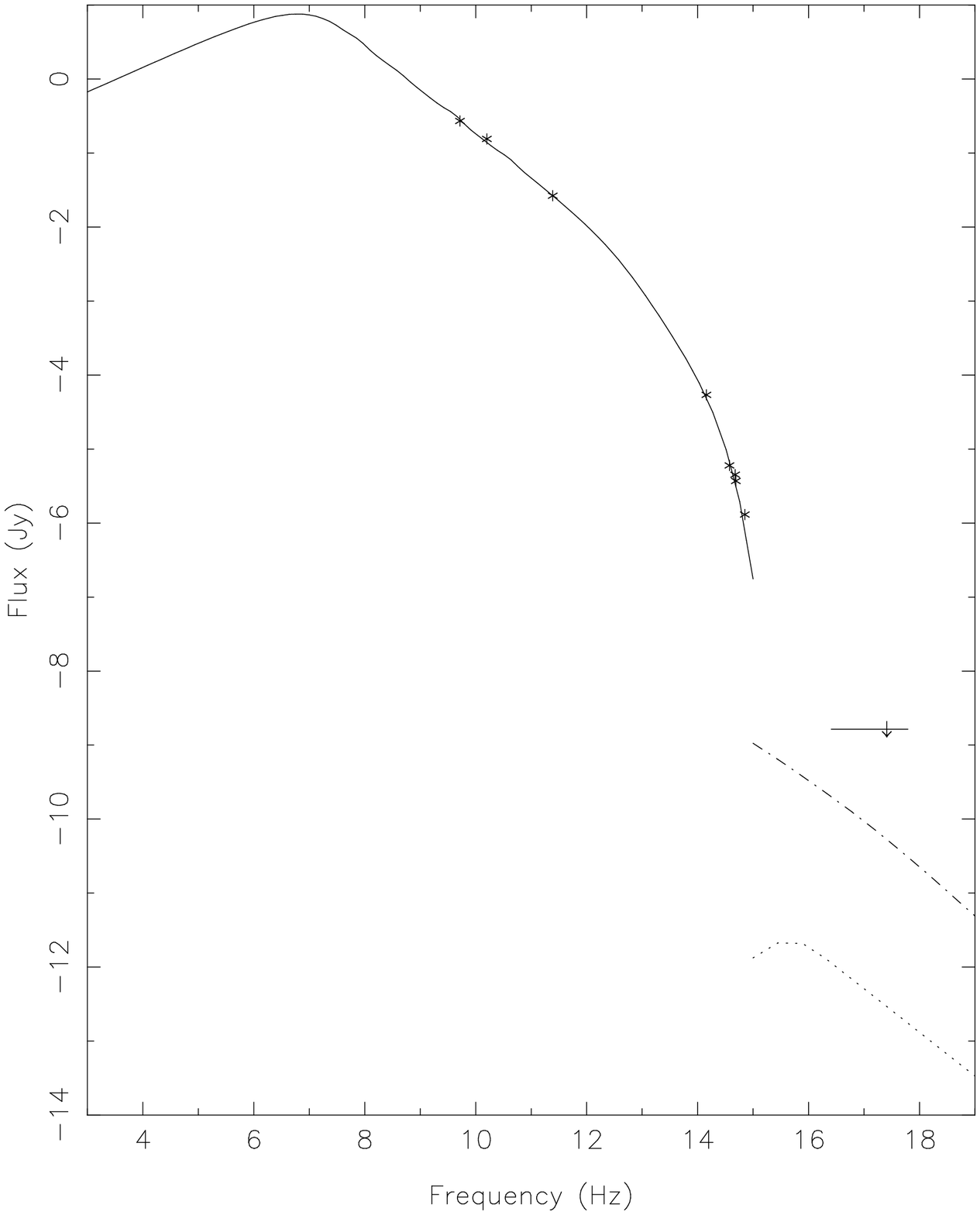}\hskip 10pt\epsfysize 10.5cm
\epsfbox{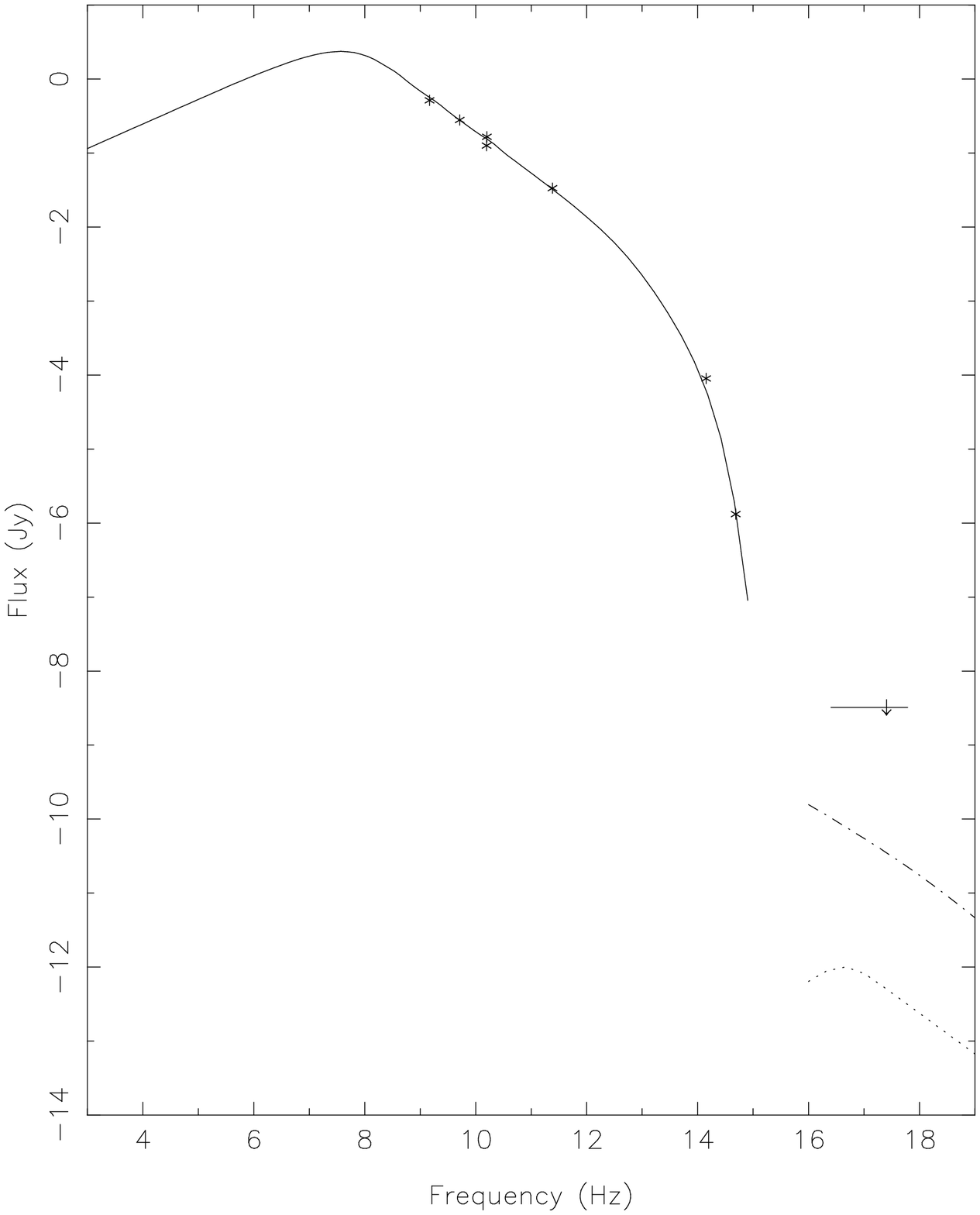}}
\caption{Spectra of the two hot spots. Left: 3C\,33 S. Right: 3C\,111
N. Scales are logarithmic to base 10. Radio and optical points (stars)
are taken from Meisenheimer \etal\ (1989); the arrows denote the X-ray
upper limit at 1 keV and the lines through them show the 0.1--2.4 HRI
energy passband. The solid lines are fitted synchrotron spectra, with
breaks and high-energy cutoffs. The dot-dashed lines show the
predicted SSC emission and the dotted lines are the predicted
contribution from scattered CMB photons.}
\label{spectra}
\end{center}
\end{figure*}

No emission is detected in the region corresponding to the radio hot
spot. The background per ($\sim 5 \times 5$ arcsec$^{2}$) detection
cell, estimated from an annulus between radii of 125 and 175 arcsec
around the central source, is $3.0 \times 10^{-5}$ counts sec$^{-1}$;
if Poisson statistics apply, a $3\sigma$ detection (i.e. one with a
probability $<0.3$ per cent of occurring by chance) would be $>5$ counts, or
$> 1.0 \times 10^{-4}$ counts s$^{-1}$. We may take this rate as an
upper limit on the X-ray emission from the undetected hot spot. The
detection cell size is chosen so that 50 per cent of the source photons fall
within a cell given the instrument PSF; we therefore multiply this
rate by 2 to get the total limiting count rate from the undetected
component. Assuming a power law spectrum for the emission with energy
spectral index 0.5, similar to that seen in the radio, and galactic
absorption, we calculate (using {\sc pimms}) that this corresponds to
a 1-keV flux density upper limit of approximately 1.6 nJy. This limit
is relatively insensitive to spectral assumptions, since 1 keV is
close to the central energy of the HRI bandpass. In Fig.\
\ref{spectra} we plot this limit together with the measured optical
and radio flux densities of the hot spot, taken from Meisenheimer
\etal\ (1989). (We use their decomposition of the hot spot radio flux
into contributions from an extended and a compact component.) It will
be seen that the X-ray upper limit lies well below any straight-line
extrapolation of the radio or radio-to-optical spectrum.

\subsection{3C111}

3C\,111 was observed with the HRI for a total of 12.9
ks between 1996 September 15 and 17. The data were reduced in a
similar way to those from 3C\,33, and after high-background time filtering
there were 10.5 ks of good data.

\begin{figure*}
\begin{center}
\leavevmode
\epsfysize 10cm
\epsfbox{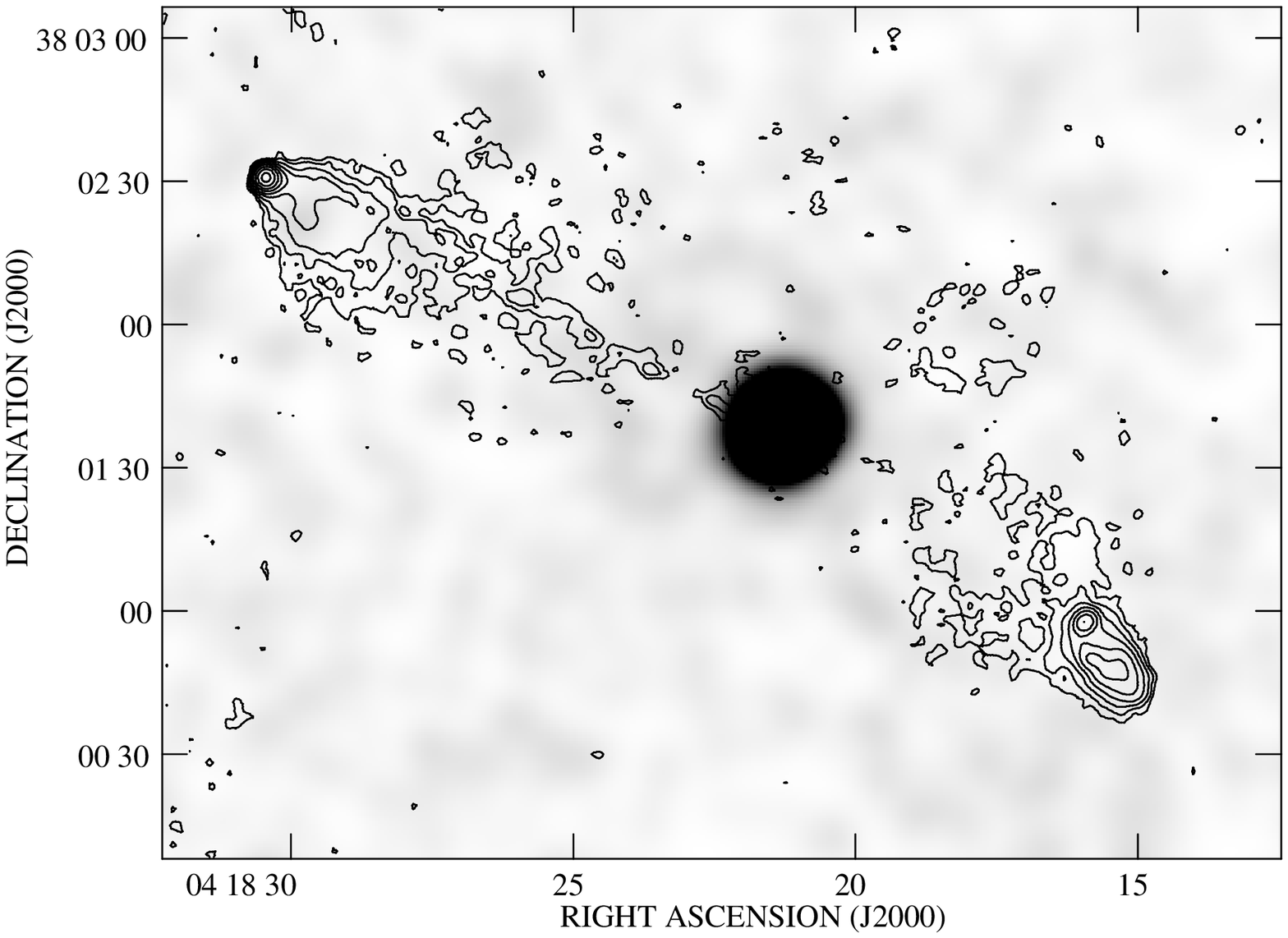}
\caption{8.4-GHz radio contours of 3C\,111 at 2.5-arcsec resolution (from Leahy \etal\ 1997) overlaid
on the HRI X-ray image, smoothed with a Gaussian with $\sigma =
4$ arcsec. Radio contours at $1 \times (1,2,4\dots)$ mJy beam$^{-1}$.}
\label{3C111-pic}
\end{center}
\end{figure*}

Fig.\ \ref{3C111-pic} shows an overlay of the radio contours of Leahy
\etal\ (1997) on the X-ray image. A strong source was detected at a
J2000 position of 04 18 21.3 +38 01 38.5, approximately 2.5 arcsec
away from the radio core position of Linfield \& Perley (1984). Using
source and background regions identical in size to those used for
3C\,33, the count rate for the source is $\sim 0.073$ s$^{-1}$, or
$760 \pm 30$ counts in total. When a broadened PSF is fitted to the
radial profile from 13 logarithmically spaced bins, as described
above, the best-fit FWHM is 4 arcsec and the 99 per cent upper limit
on the FWHM is 6 arcsec. The fit in this case is poorer than for
3C\,33, suggesting either that some real extension is present or that
the aspect-error extension is poorly modelled by a simple broadened
PSF. Nevertheless we can be confident that most of the emission comes
from a region less than 8 kpc in size.

3C\,111 has been observed in the X-ray with a number of other
instruments. It was detected with EXOSAT (Turner \& Pounds 1991), with
the {\it Einstein} IPC (Wilkes \etal\ 1994), with {\it Ginga} (Nandra
\& Pounds 1994) and with the {\it Einstein} SSS and {\it HEAO-1}
(Turner \etal\ 1991). Spectral fits from the literature are tabulated
by Malaguti, Bassani \& Caroli (1994). There are two {\it ROSAT} PSPC
observations in the public archive, one of 0.89 ks taken on 1991 Mar
08 and one of 2.32 ks taken on 1993 Feb 13.

The X-ray spectrum of the source appears to be strongly variable, with
both absorbing columns and photon indices being inconsistent at
different epoch and with different instruments (e.g.\ Turner \etal\
1991). In an attempt to provide a recent soft X-ray measurement, we
used PROS to fit power-law models to the later of the archival PSPC
datasets. Allowing energy index and intrinsic $N_H$ to vary
independently, we find a best-fit $N_H = 5_{-4}^{+5} \times 10^{21}$
cm$^{-2}$, where the errors are $1\sigma$ for two interesting
parameters; the power-law photon index $\Gamma$ is poorly constrained
by the data, with a best-fit $\Gamma = 1.3 \pm 1.2$. These fits are
consistent with galactic absorption (galactic $N_H = 3.26 \times
10^{21}$ cm$^{-2}$, Elvis \etal\ 1989). The photon index is lower
than, though not inconsistent with, soft X-ray indices from the {\it
Einstein} IPC observations taken in 1980 ($\Gamma =
2.25_{-1.75}^{+1.95}$).

The broad-band (0.1-2.4 keV) count rates derived from the earlier and
later PSPC observations are $0.28 \pm 0.03$ and $0.42 \pm 0.01$ counts
s$^{-1}$, which are significantly inconsistent with one another. Using
the best-fit spectral models, the later PSPC observation would imply a
count rate with the HRI of 0.15 counts s$^{-1}$, approximately twice
the count rate we derive from the present HRI data. Because of the
very similar spectral responses of the HRI and PSPC, this result is
only weakly affected by spectral assumptions; the {\it ROSAT}
observations thus show that luminosity in the soft X-ray band varies
by about a factor of two on timescales of two years. The radio core is
also variable, with changes of $\sim 10$ per cent in a few
months (Leahy \etal\ 1997).

The later PSPC observation implies an 0.1--2.4 keV luminosity of $3
\times 10^{44}$ ergs s$^{-1}$ for 3C\,111, roughly 50 times larger
than the luminosity of 3C\,33 if the X-rays from that object are not
seen through a large intrinsic column density. The sources are very
similar in their large-scale radio properties, but 3C\,111 is a
broad-line and 3C\,33 a narrow-line radio galaxy. In unified models
for radio sources, broad-line objects are expected both to suffer less
from obscuration and to have any relativistically moving component
enhanced by beaming. 3C\,111 is 50 times more luminous in the soft
X-ray, and the ratio of radio core luminosities in the two sources is
also approximately 50. Given that the radio core emission is expected
to originate outside the region of strong absorption, and that the
PSPC spectrum of 3C\,111 is consistent with galactic absorption, this
might be taken as an indication that the soft X-ray emission in these
sources is dominated by a component directly related to the radio
emission, for example by the SSC process. However, the high and
apparently variable absorption measured in the hard X-ray band for
3C\,111 indicates that this is not the whole story. A combination of a
soft unabsorbed component and a hard absorbed component, both
variable, may be needed to explain the broad-band X-ray spectral
properties of the source.

Again, no emission is detected from the region corresponding to the
radio hot spot; the background per detection cell, using the same
annulus, is $3 \times 10^{-5}$ counts s$^{-1}$, giving an upper
limit of 2 counts in the cell over the time interval. This corresponds
to a flux upper limit of 3.2 nJy.
In Fig.\ \ref{spectra} we plot
this limit together with the radio and optical flux densities from
Meisenheimer \etal\ (1989).

\section{Discussion}

The fact that the upper limits on hot spot X-ray flux density fall
below any plausible straight-line extrapolation of the radio (or
radio-to-optical) spectrum gives us an independent reason for
confidence in the energy cutoffs in the electron spectrum inferred by
Meisenheimer \etal\ (1989). If synchrotron emission is insignificant
at X-ray frequencies, we may use the upper limits obtained as a
constraint on the flux from inverse-Compton scattering. The hot spots
may be modelled as isotropic spheres (errors arising from the
assumption of this simple geometry are of order unity) with a volume
equal to that quoted by Meisenheimer \etal . The radio-to-optical
synchrotron spectra are best fitted by a power-law electron energy
spectrum with a high-energy spectral break ($N(E)$ proportional to
$E^{-p}$ below $E_{\rm break}$ and to $E^{-(p+1)}$ afterwards); we
consider this energy spectrum as being bounded by low and high energy
rather than frequency cutoffs (e.g.\ Myers \& Spangler 1985). Input
and fitted parameters are listed in Table \ref{fitpar}; with a broken
spectrum, the dependence of the results on the high energy cutoff
($E_{\rm max}$) is negligible. With a minimum electron energy given by
$E_{\rm min} = \gamma_{\rm min}m_{\rm e}c^2$, with $\gamma_{\rm min} =
100$, we find best-fit equipartition magnetic fields (with no protons)
of $B_{\rm eq}$ = 30 nT for 3C\,33 and 21 nT for 3C\,111.

Once the $B$-field is
known we may use the simple analysis of Rybicki \& Lightman (1979),
chapter 7, to calculate the expected flux in X-rays from the SSC
process and from inverse-Compton scattering of other photon sources,
using the relationship of Band \& Grindlay (1985) between the
emissivity and the average photon number density in an isotropic
sphere. The effects of multiple scatterings and of the high-energy
(Klein-Nishina) corrections to the Thomson cross-section can both be
neglected in the energy and density regime of interest. The
synchrotron-self-Compton emissivity at a given frequency ($J_{\rm
ic}(\nu_1)$, in W Hz$^{-1}$ m$^{-3}$) is then given by
\begin{displaymath}
J_{\rm ic}(\nu_1) = {9\over 16} m_{\rm e}^2 c^4\nu_1\sigma_{\rm T} R\int^{E_{\rm max}}_{E_{\rm min}} \int^{\nu_{\rm max}}_{\nu_{\rm min}}{{N(E) J(\nu_0)}\over{E^2
\nu^2_0}} f(x) {\rm d}\nu_0 {\rm d}E 
\end{displaymath}
where $m_e$ is the electron mass, $c$ is the speed of light,
$\sigma_T$ is the Thomson cross-section, $R$ is the radius of the
sphere, $N(E)$ is the number density of electrons as a function of
electron energy, $J(\nu_0)$ is the synchrotron emissivity as a
function of frequency, and $f(x)$ is a function of $E$, $\nu_1$ and
$\nu_0$ defined by Rybicki \& Lightman. The integration can then be performed
numerically for an arbitrary electron energy spectrum.

The results of these calculation are plotted in
Fig.\ \ref{spectra}. It will be seen that the
X-ray upper limits lie more than an order of magnitude above the
predicted SSC flux at equipartition, with the contribution from
inverse-Compton scattering of the CBR being much smaller. At
distances of hundreds of kpc from the nucleus, the energy density in
photons originating at an AGN with bolometric power $\sim 10^{39}$ W
is expected to be smaller than that in the CBR, and so we ignore the
contribution to inverse-Compton flux from this
source.\footnote{Brunetti, Setti \& Comastri (1997) discuss the
possibility that inverse-Compton scattering from the lobes contributes
significantly to the X-ray fluxes of distant powerful radio galaxies.
In the objects considered here, we see no evidence for the {\it
extended} X-ray emission that would be produced by this
process. However, the electron densities, and possibly also the hidden
quasar luminosities, are lower in our objects.} 

The flux expected from the SSC process depends on the number density
of electrons, so variations in the volume of the hot spot or in the
magnetic field (given that a fixed emissivity must be produced to fit
the observed radio spectrum) cause it to change; we can use this to
investigate the physical conditions in the hot spot. The results are
insensitive to the precise position of the electron energy break or to
the upper energy cutoff. The dependence on the volume of the hot spot
is also weak, but it allows us to rule out radii for the hot spots of
less than a few parsecs if equipartition holds; these are of course
much less than the sizes implied by radio imaging. More interesting
are the constraints on magnetic field strength; with the hot spot
volume as given by Meisenheimer \etal\ (1989), the magnetic field in
the hot spot of 3C\,33 cannot be less than about 4 nT, or about 7
times weaker than equipartition, with $\gamma_{\rm min} = 100$. The
relationships between the limiting $B$-field value, $B_{\rm eq}$ and
$\gamma_{\rm min}$ for the two sources are shown graphically in Fig.\
\ref{exclude}. It will be seen that a high $\gamma_{\rm min}$ weakens
the constraints on $B$. $\gamma_{\rm min}$ is only weakly constrained
($\gamma_{\rm min} \la 2000$) by the observations in the cases
considered here, but the spectral shape of the hot spots in Cygnus A
led Carilli \etal\ (1991) to suggest $\gamma_{\rm min} \sim 400$ in
that object.

If there is a significant contribution from relativistic protons
(expressed as a ratio $\kappa$ of energy density in non-radiating
particles to energy density in radiating particles) or a plasma
filling factor $\phi$ less than unity, the predicted SSC flux falls,
since the $B$-field increases to maintain equipartition with the
increased energy density in relativistic particles and the number
density of electrons must fall to maintain the synchrotron emissivity
at the correct level. Thus with $\kappa = 10$ ($\phi = 1$) our fitted
equipartition field in the hot spot ($\gamma_{\rm min}=100$) is 59 nT,
and the SSC flux is nearly two orders of magnitude below the X-ray
limit. In principle, therefore, observations which measure SSC
emission would allow us to distinguish between models with different
values of the parameter $(1+\kappa)/\phi$, if some independent
estimator of the magnetic field were available. Large $\kappa$ or
small $\phi$ allow the sources to be further out of equipartition
without violating the constraints from X-ray observations.

\section{Conclusions}

The {\it ROSAT} HRI has detected a weak X-ray source coincident with
the radio core of 3C\,33, and has made a good soft-X-ray measurement
of the nuclear X-ray source of 3C\,111. The relative powers of the two
sources are in agreement with the trend seen in other work
(e.g.\ Siebert \etal\ 1996) for narrow-line radio galaxies to be much
fainter in soft X-ray than their broad-line counterparts. Comparison
with earlier data provides further evidence for X-ray variability in the
core of 3C\,111. The soft X-ray emission in both sources can plausibly
be associated with unabsorbed emission from the radio cores.

The hot spots of both sources are undetected in the X-ray. This allows
us to rule out any plausible extension of the synchrotron spectrum to
this high frequency, which differs from the synchrotron explanation
proposed for 3C\,390.3 and for the jets in M87 and 3C\,273. Because
SSC emission from the hot spots is inevitable, the magnetic fields in
the hot spots cannot be more than an order of magnitude less than
equipartition for $\kappa = 0$, $\phi=1$.

X-ray detections of SSC emission from hot spots would allow a
relatively model-independent measurement of the magnetic field. The
{\it ROSAT} HRI is not an ideal instrument for this purpose, as its
PSF smears out the flux from compact objects such as hot
spots. {\it AXAF} should allow the detection of SSC emission from the
hot spots of radio sources such as those described here.

\section*{ACKNOWLEDGEMENTS}

We are grateful to Paddy Leahy for providing the radio image of 3C\,33
and to Dan Harris for discussions. We thank an anonymous referee for a
number of comments that enabled us to improve the clarity of the
paper. This research has made use of the NASA/IPAC Extragalactic
Database (NED) which is operated by the Jet Propulsion Laboratory,
California Institute of Technology, under contract with the National
Aeronautics and Space Administration. Support from PPARC grant
GR/K98582 and NASA grant NAG 5-2312 is gratefully acknowledged.

\bsp

\begin{table}
\caption{The radio properties of 3C\,33 and 3C\,111}
\label{sources}
\begin{tabular}{llll}
Source&$z$&178-MHz luminosity&Total radio extent\\
&&(W Hz$^{-1}$ sr$^{-1}$)&(kpc)\\
3C\,33 &0.0595&$7.5\times 10^{25}$&400\\
3C\,111&0.0485&$5.9\times 10^{25}$&280\\
\end{tabular}
\vbox{\medskip
Redshifts and radio fluxes are taken from Spinrad \etal\ (1985).}
\end{table}

\begin{table*}
\caption{Best-fit parameters of the hot spots assuming equipartition. Calculated with $\gamma_{\rm min} =
100$ so that $E_{\rm min} =51$ MeV, no protons ($\kappa=0$) and
filling factor unity ($\phi=1$).}
\label{fitpar}
\begin{tabular}{llllllll}
Hot spot&Adopted radius&Low-energy electron&Fitted $B_{\rm eq}$&$E_{\rm
max}$&$E_{\rm break}$&Predicted SSC flux&Observed
1-keV flux\\
&(kpc)&power law index ($p$)&(nT)&(GeV)&(GeV)&density (nJy)&density (nJy)\\
3C\,33 S&0.27&2.18&30&220&28&0.059&$<1.6$\\
3C\,111 N&0.37&2.10&21&210&44&0.041&$<3.2$\\
\end{tabular}
\end{table*}

\newlength{\halfwidth}
\setlength{\halfwidth}{\linewidth}
\begin{figure*}
\begin{center}
\leavevmode
\hbox{
\epsfxsize=\halfwidth
\epsfbox{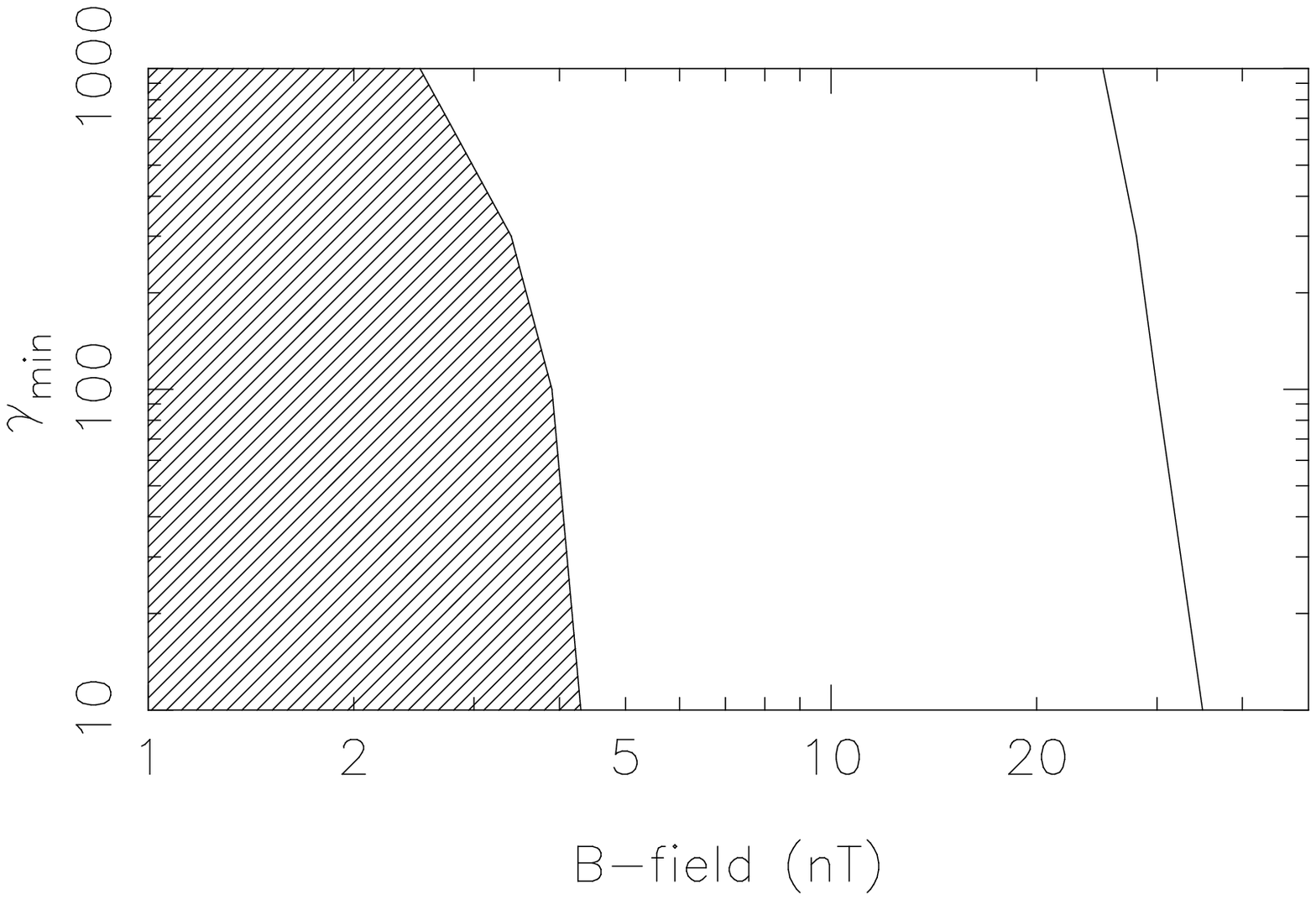}
\epsfxsize=\halfwidth
\epsfbox{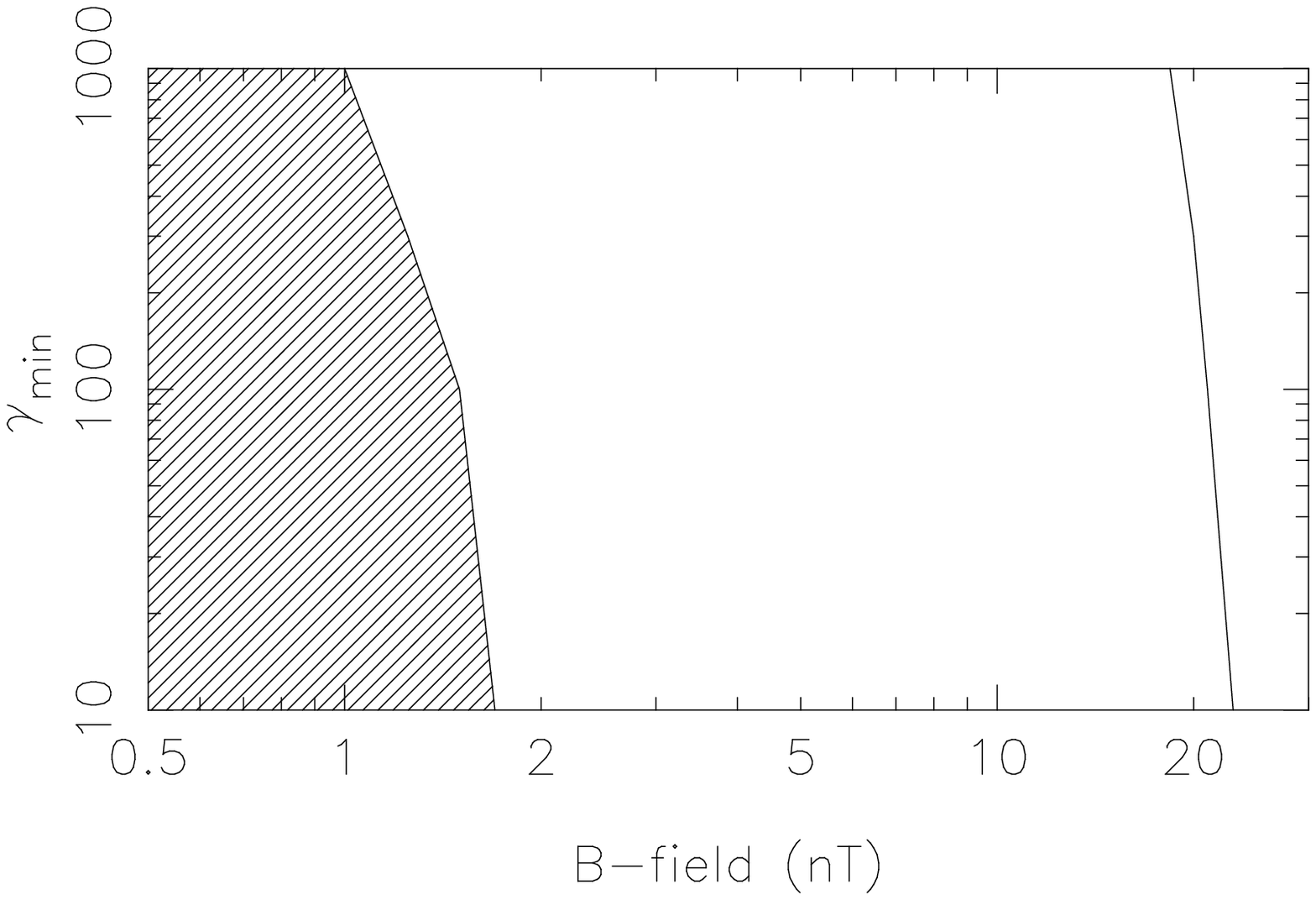}
}
\caption{Regions of parameter space excluded by the limits on X-ray
emission. Left: 3C\,33. Right: 3C\,111. The shaded area shows the
excluded regions, while the solid line shows the calculated
equipartition field (with $\kappa=0$, $\phi=1$).}
\label{exclude}
\end{center}
\end{figure*}

\end{document}